\documentclass[preprint,amsmath,amssymb,prd,superscriptaddress]{revtex4}
\usepackage[dvips]{graphicx}
\usepackage{color}
\usepackage{multirow}
\usepackage{array}
\usepackage{float}
\usepackage{amsmath,amssymb,amsfonts,amstext,amsthm}
\usepackage{algorithm,algorithmic}
\usepackage{kpfonts}
\makeatletter
\newcommand{\doublewidetilde}[1]{{%
  \mathpalette\double@widetilde{#1}%
}}
\newcommand{\double@widetilde}[2]{%
  \sbox\z@{$\m@th#1\widetilde{#2}$}%
  \ht\z@=.9\ht\z@
  \widetilde{\box\z@}%
}
\makeatother
\newcommand{\p}{\partial}
\newcommand{\pslash}{p\kern-1ex /}
\newcommand{\lslash}{l\kern-1ex /}
\newcommand{\kslash}{k\kern-1ex /}
\newcommand{\dslash}{\p\kern-1.2ex /}
\newcommand{\Dslash}{{\cal D}\kern-1.5ex /}
\newcommand{\Tr}{{\rm Tr}}
\newcommand{\tr}{{\rm tr}}
\newcommand{\re}{{\rm Re}}

\newcommand{\diag}{{\rm diag}}

\newcommand{\bea}{\begin{eqnarray}}
\newcommand{\eea}{\end{eqnarray}}

\newcommand{\BAN}{\begin{eqnarray*}}
\newcommand{\EAN}{\end{eqnarray*}}

\begin{document}

\newcommand{\NTNU}{
  Physics Department, National Taiwan Normal University, Taipei, Taiwan~11677, R.O.C.
}

\newcommand{\ASIOP}{
  Institute of Physics, Academia Sinica, Taipei, Taiwan~11529, R.O.C.
}

\newcommand{\NCTS}{
  Physics Division, National Center for Theoretical Sciences, 
  National Tsing-Hua University, Hsinchu, Taiwan~30013, R.O.C.
}

\newcommand{\NTU}{
  Physics Department, National Taiwan University, Taipei, Taiwan~10617, R.O.C.
}

\preprint{NTUTH-18-505A}

\title{Improved study of the $\beta$-function of $SU(3)$ gauge theory with $N_f = 10 $ massless 
       domain-wall fermions}

\author{Ting-Wai~Chiu}
\affiliation{\NTNU}
\affiliation{\ASIOP}
\affiliation{\NTU}

\begin{abstract} 

I perform an improved study of the $\beta$-function of $ SU(3) $ lattice  
gauge theory with $N_f=10$ massless optimal domain-wall fermions in the fundamental representation,  
which serves as a check to what extent the scenario 
in the previous work [arXiv:1603.08854; Proc. Sci. LATTICE2016 (2017) 228] is valid.
In the finite-volume gradient flow scheme with $ c = \sqrt{8t}/L = 0.3 $, 
the renormalized couplings $g^2 (L,a) $ of four primary lattices ($ L/a = \{ 8, 10, 12, 16 \}$) are 
tuned (in $ 6/g_0^2 $) to the same $ g_c^2 $ with statistical error less than $0.5 \% $,  
in contrast to the previous work
where $ g^2(L,a) $ were obtained by the cubic-spline interpolation.
Then the renormalized couplings $ g^2(sL, a) $ of the scaled lattices ($ sL/a = \{16, 20, 24, 32\} $
with $s=2$) are computed at the same $ 6/g_0^2 $ of the corresponding primary lattices.   
Using the renormalized couplings of four lattice pairs 
$ (sL,L)/a = \{ (16,8), (20,10), (24,12), (32,16) \} $, 
the step-scaling $\beta$-function $ [g^2(sL,a) - g^2(L,a)]/\ln (s^2) $ is computed and  
extrapolated to the continuum limit $ \beta(s,g_c^2) $, as summarized in Table \ref{tab:DBF_a0}. 
Based on the four data points of $ \beta(s,g_c^2) $ at   
$ g_c^2 = \{ 6.86(2), \ 6.92(3), \ 7.03(2), \ 7.16(2) \} $, I infer that the theory
is infrared near-conformal, or conformal with the fixed-point $ g_*^2 = 7.55(36) $. 
This corrects the scenario in the previous work with $ g_*^2 \sim 7.0 $,  
and also suggests that the interpolation method cannot give a reliable   
determination of the $\beta$-function, especially in the regime close to the infrared fixed-point.

\end{abstract}

\maketitle


\section{Introduction}

In Refs. \cite{Chiu:2016uui,Chiu:2017kza}, I investigated the $\beta$-function 
of the $SU(3)$ gauge theory with $N_f=10$ massless optimal domain-wall fermions
in the fundamental representation. The motivation was to see whether this theory
possesses a non-trivial infrared fixed point, 
which is not only a fundamental problem in quantum field theory, 
but also relevant to beyond the Standard Model scenarios with a composite Higgs boson. 
(For recent reviews, see, e.g., Refs. \cite{Pica:2017gcb, Svetitsky:2017xqk, Witzel:2018abc}.)
The results in Refs. \cite{Chiu:2016uui,Chiu:2017kza} suggest that 
the theory might possess an infrared fixed point (IRFP) around $ g_c^2 \sim 7 $. 
However, the major systematic uncertainty in Refs. \cite{Chiu:2016uui,Chiu:2017kza} 
was that interpolation was used 
to obtain the renormalized couplings $ g^2(L,a) $ and $ g^2(sL, a) $. This could lead to  
a large systematic error in the strong-coupling regime where the renormalized coupling 
varies rapidly with respect to the bare coupling $ g_0 $ (or $ 6/g_0^2 $), 
which in turn may give incorrect results for the step-scaling $\beta$-function 
\bea
\label{eq:DBF}
\beta(s,a/L, g^2) = \frac{g^2(sL,a) - g^2(L,a)}{\ln (s^2)},  
\eea
as well as its extrapolated value in the continuum limit ($ a/L \to 0$). 
The purpose of the present study is to eliminate this systematic uncertainty, 
by tuning $ 6/g_0^2 $ such that the renormalized couplings $g^2(L,a)$ of 
all primary lattices ($ L/a = 8, 10, 12, 16 $) have 
the same value with statistical error less than $0.5 \% $. 
The tuning process implies that many simulations on the primary lattices 
have to be performed, which are rather challenging 
in terms of computing resources, time, and effort.
After the value of $ 6/g_0^2 $ is determined for a chosen $ g_c^2 = g^2(L,a) $,  
the simulation on the scaled ($s=2$) lattice is performed at the same $ 6/g_0^2 $ 
to obtain the renormalized coupling $ g^2(sL,a) $. 
Since the results in Refs. \cite{Chiu:2016uui,Chiu:2017kza} suggest that 
the theory may possess an infrared fixed point around $ g_c^2 \sim 7 $, 
four targeted values of $ g_c^2 = \{ 6.86(2), \ 6.92(3), \ 7.03(2), \ 7.16(2) \} $ around $ g_c^2 = 7.0 $ 
are chosen. Also, a point at $ g_c^2 = 3.51(2) $ is picked to check whether the 
interpolation used in Refs. \cite{Chiu:2016uui,Chiu:2017kza} works well in the regime 
where the renormalized coupling varies slowly with respect to the bare coupling. 
Moreover, in view of a recent study of the $SU(3)$ gauge theory with $ N_f = 10 $ massless 
domain-wall fermions \cite{Hasenfratz:2017qyr} (which reported 2-3 standard deviations 
compared with the results of Ref. \cite{Chiu:2017kza} for $ 4.5 < g_c^2 < 6.0 $),  
a point at $ g_c^2 = 5.25(2) $ is chosen to check whether the discrepancy is due to the 
systematic error of the interpolation used in Ref. \cite{Chiu:2017kza}.   
All together, the targeted values of $ g_c^2 $ in this study are 
$ g_c^2 = \{ 3.51(2), \ 5.25(2),\ 6.86(2), \ 6.92(3), \ 7.03(2), \ 7.16(2) \} $.

The outline of this paper is as follows. 
In Section II, we describe our hybrid Monte Carlo (HMC) simulation of $SU(3)$ 
gauge theory with $ N_f = 10 $ massless 
optimal domain-wall fermions, and summarize the residual masses of all gauge ensembles 
in Table \ref{tab:mres}. In Section III, we present our results for the renormalized couplings in      
the finite-volume gradient flow scheme with $ c = \sqrt{8t}/L = 0.3 $, for all gauge ensembles, 
as summarized in Table \ref{tab:g2L}. In Section IV, we perform the extrapolation 
of the step-scaling $\beta$-function $ \beta(s,a/L,g_c^2) $ to the continuum limit ($ a/L \to 0 $) 
with the linear fit $ [A + B (a/L)^2 ]$ and quadratic fit [$ A + B (a/L)^2 + C(a/L)^4 $]. 
The results are summarized in Table \ref{tab:DBF_a0}.    
In Section V, we perform the extrapolation of $ \beta(s,g_c^2) $ with the linear fit, 
using four data points at $ g_c^2 = \{ 6.86(2), \ 6.92(3), \ 7.03(2), \ 7.16(2) \} $, 
and determine the IRFP $ g_*^2 $ and the slope of $ \beta(s,g^2) $ at the IRFP. 
In Section VI, we determine the universal scaling exponent $ \gamma_g^* $ of the conventional 
$\beta$-function $\beta(g^2(\mu)) $ in the continuum, with the input of   
the slope of $ \beta(s,g^2) $ at the IRFP. 
In Section VII, we summarize the results of this paper,   
and discuss the discrepancies between the results in this paper and those  
obtained with $ N_f = 10 $ massless staggered fermions in a recent study 
\cite{Nogradi:2018abc}.

\section{Generation of the Gauge Ensembles}

Since we are dealing with massless fermions, it is vital to use lattice fermions with 
exact chiral symmetry at finite lattice spacing 
(i.e., domain-wall \cite{Kaplan:1992bt} /overlap \cite{Neuberger:1997fp} fermions) 
with exactly the same flavor symmetry as their counterpart in the continuum.
Theoretically, the effective four-dimensional lattice Dirac operator of the domain-wall fermion 
with infinite extent in the fifth dimension ($N_s = \infty $) is exactly equal to the
overlap Dirac operator  
\BAN
D(m_q) = m_q + \frac{(1 - r m_q )}{2r} \left[ 1 + \gamma_5 H (H^2)^{-1/2} \right], 
\EAN
where $ m_q $ is the bare fermion mass, 
\BAN 
r &=& \frac{1}{2m_0(1-d m_0)}, \hspace{4mm} m_0 \in (0,2), \\
H &=& c \gamma_5 D_w ( 1 + d D_w )^{-1}, 
\EAN  
$ D_w $ is the standard Wilson-Dirac operator minus $m_0$, 
and $ c $ and $ d $ are parameters depending on the variant of the domain-wall fermions. 
In this study, we set $ c = 1 $ and $ d = 0 $, and thus $ H = \gamma_5 D_w = H_w $. 
In practice, the sign function $ S(H) \equiv H(H^2)^{-1/2} $ cannot be computed exactly, since $ H $ is 
a very large matrix and it is prohibitively expensive to diagonalize $ H $. 
The best way to proceed is to use the Zolotarev optimal rational approximation of the sign function 
$ S(H) $. However, HMC \cite{Duane:1987de} simulations on the four-dimensional lattice 
with the overlap-Dirac operator in the Zolotarev approximation encounter enormous difficulties 
(see, e.g., Refs. \cite{Fodor:2003bh,DeGrand:2006ws}).
On the other hand, for domain-wall fermions (DWFs) with finite $ N_s $ in the fifth dimension, the 
HMC simulations can be performed without serious difficulties. However, DWFs could severely break 
the exact chiral symmetry, depending on the approximate sign function $ S(H) $ 
in the four-dimensional effective Dirac operator.  

The chiral symmetry can be maximally preserved on a lattice
with finite $N_s$ by optimal domain-wall fermions \cite{Chiu:2002ir}, 
with the effective four-dimensional lattice Dirac operator exactly equal to the Zolotarev 
optimal rational approximation of the overlap Dirac operator. 
In this paper, we use optimal DWFs with the $ R_5 $ symmetry \cite{Chiu:2015sea},  
whose effective four-dimensional lattice Dirac operator exactly 
equal to the ``shifted" Zolotarev optimal rational approximation 
of the overlap operator, with the approximate sign function $ S(H) $ 
satisfying the bound $ 0 \le 1-S(\lambda) \le 2 d_Z $ 
for $ \lambda^2 \in [\lambda_{min}^2, \lambda_{max}^2] $,
where $ d_Z $ is the maximum deviation $ | 1- \sqrt{x} R_Z(x) |_{\rm max} $ of the
Zolotarev optimal rational polynomial $ R_Z(x) $ of $ 1/\sqrt{x} $ 
for $ x \in [1, \lambda_{max}^2/\lambda_{min}^2] $, with degrees $ (n-1,n) $ for $ N_s = 2n $.

The action of one-flavor optimal DWFs can be written as   
\bea
\label{eq:S_odwf}
S(\bar\Psi, \Psi, U) = \bar\Psi_{x,s} \left[
  (\omega_s D_w + 1)_{xx'} \delta_{ss'}
 +(\omega_s D_w - 1)_{xx'} L_{ss'} \right] \Psi_{x',s'},  
\eea
where the indices $ x $ and $ x' $ denote the sites on the four-dimensional spacetime lattice,
$ s $ and $ s' $ are the indices in the fifth dimension, and 
the lattice spacing $ a $ and the Dirac and color indices have been suppressed.
Here $ D_w $ is the standard Wilson-Dirac operator minus the parameter $ m_0 \in (0,2) $.  
The operator $ L $ is independent of the gauge field, and it can be written as  
\BAN
L = P_+ L_+ + P_- L_-, \quad P_\pm = (1 \pm \gamma_5)/2,
\EAN
and
\bea
\label{eq:L}
(L_+)_{ss'} = (L_-)_{s's} = \left\{ 
    \begin{array}{ll} 
      - m_q/(2 m_0) \ \delta_{N_s,s'}, & s = 1, \\  
          \delta_{s-1,s'}, & 1 < s \leq N_s,   
    \end{array}\right.
\eea
where $ m_q $ is the bare fermion mass, $ m_0 \in (0, 2) $,  
and $ N_s $ is the number of sites in the fifth dimension,    
For massless DWFs, $ m_q $ is set to zero.  
Besides Eq. (\ref{eq:S_odwf}),    
the action for the Pauli-Villars fields with $ m_q = 2 m_0 $ has to be included   
for the cancellation of the bulk modes, which is exactly the same as Eq. (\ref{eq:S_odwf}) 
except for $ m_q = 2 m_0 $ in $ L_\pm $ [Eq. \ref{eq:L}]. 
Thus the action for $ SU(3) $ lattice gauge theory with $N_f=10$ massless optimal DWFs can be written as 
\BAN
S_g(U) + \sum_{f=1}^{10} \left\{ S_{m_q=0}( \bar\Psi, \Psi, U)_f + S^{PV}_{m_q=2 m_0} (\bar\Phi, \Phi, U)_f \right\}. 
\EAN
where $ S_g(U) $ is the gauge action. In this paper, we use the Wilson plaquette gauge action 
\BAN
S_g(U) = \frac{6}{g_0^2} \sum_{plaq.}\left\{1-\frac{1}{3} \re \Tr (U_p) \right\}, 
\EAN
where $ g_0 $ is the bare coupling.
For the fermion action, we set $ m_0 = 1.8 $, and $ N_s = 16 $.
The optimal weights $ \omega_s $  \cite{Chiu:2015sea} 
are computed with $ \lambda_{max}/\lambda_{min} = 6.2/0.05 $. 

Simulating $ N_f = 10 $ DWFs amounts to simulating five pairs of $ N_f = 2 $ DWFs.
Starting from the action (\ref{eq:S_odwf}) and   
following the procedures of even-odd preconditioning 
and the Schur decomposition given in Ref. \cite{Chiu:2013aaa}, 
the partition function for the $ SU(3) $ gauge theory 
with $ N_f = 10 $ massless optimal DWFs can be written as 
\bea
\label{eq:Z_nf10}
Z = \int[dU]\prod_{i=1}^5 [d\phi^{\dag}]_i[d\phi]_i 
\exp \left(-S_g[U]- \sum_{i=1}^5 \phi^\dagger_i (C_{PV}^\dagger)_i ( C C^\dagger)^{-1}_i (C_{PV})_i \phi_i \right),
\eea  
where $ \phi_i $ and $ \phi^\dagger_i $ are pseudofermion fields, and  
\BAN
\label{eq:C_def}
C &=& 1 - M_5 D_w^{\text{OE}} M_5 D_w^{\text{EO}}, \\
\label{eq:m5}
M_5 &=& \left\{(4-m_0) + \omega^{-1/2}_s [(1-L)(1+L)^{-1}]_{s,s'} \omega^{-1/2}_{s'} \right\}^{-1}.
\EAN

However, HMC simulations with Eq. (\ref{eq:Z_nf10}) turn out to be rather
time consuming for large lattices at strong couplings, e.g., $ 32^4 $ at $ 6/g_0^2 = 6.45 $. 
To resolve this difficulty, 
we use a novel $ N_f = 2 $ pseudofermion action 
based on the exact pseudofermion action for one-flavor DWFs \cite{Chen:2014hyy},  
which turns out to be more efficient than (\ref{eq:Z_nf10}). 
This novel $ N_f = 2 $ pseudofermion action for the optimal DWFs  
can be written as $ S = \Phi^\dagger K(m)^\dagger K(m) \Phi $, where 
\BAN
K(m) = 1 + \left(\frac{1-m}{1+m}\right) 
       \gamma_5 v^{T}\omega^{-1/2}\frac{1}{H_T(m)}\omega^{-1/2}v, \hspace{4mm} m = \frac{m_q}{2 m_0}, 
\hspace{4mm}
v =
\left(
\begin{array}{cc}
   v_+  &     0 \\
   0    &     v_-
\end{array}
\right)_{Dirac}.
\EAN
Here $ \omega = \diag (\omega_1, \omega_2 \cdots, \omega_{N_s})$, 
$ v_+^{T} = (-1, 1, \cdots, (-1)^{N_s})$, $ v_{-} = -v_+ $,   
$ H_T(m) = R_5 \gamma_5 D_T(m) $, and 
\BAN 
D_{T}(m) =  D_w + \omega^{-1/2} (1-L)(1+L)^{-1} \omega^{-1/2}. 
\EAN
Note that $ K(m) $ is defined on the four-dimensional lattice, 
while $ H_T(m) $ is a Hermitian operator defined on the five-dimensional lattice. 
The general form of the novel two-flavors pseudofermion action for domain-wall fermions 
with $ H = c \gamma_5 D_w(1+ d D_w)^{-1} $ will be presented in a forthcoming paper \cite{Chen:2018xyz}.

Then, the partition function for the $ SU(3) $ gauge theory with $ N_f = 10 $ 
massless optimal DWFs can be written as 
\bea
\label{eq:Z1_nf10}
Z' = \int[dU]\prod_{i=1}^5 [d\phi^{\dag}]_i[d\phi]_i 
\exp \left(-S_g[U]- \sum_{i=1}^5 \phi^\dagger_i K(0)^\dagger K(0) \phi_i \right).
\eea 
We perform HMC simulations of all gauge ensembles with Eq. (\ref{eq:Z1_nf10}) 
on the five-dimensional lattice $ L^4 \times 16 $, for $ L/a = \{8, 10, 12, 16, 20, 24, 32 \} $. 
The boundary conditions of the gauge field are periodic in all directions, 
while the boundary conditions of the pseudofermion fields are antiperiodic in all directions. 
In the molecular dynamics, we use the Omelyan integrator \cite{Omelyan:2001},
and the Sexton-Weingarten multiple-time scale method \cite{Sexton:1992nu}.
Moreover, we introduce an auxiliary heavy fermion field with mass $ m_H a = 0.1 $ 
($ m_q \ll m_H \ll m_{PV} $) similar to the case of the Wilson fermion \cite{Hasenbusch:2001ne}, 
the so-called mass preconditioning.
The simulations are performed on GPU clusters with Nvidia GPUs 
(P100, GTX-1080Ti, GTX-1080, GTX-1070, GTX-1060, GTX-TITAN-X, GTX-TITAN-Z, and GTX-TITAN). 
Thermalization of each ensemble is performed on one computing node with $1-2$ GPUs. 
Then, the thermalized configurations are distributed to $16-32$ nodes 
for independent HMC simulations in multiple streams.     
For each gauge ensemble, we generate $4000-20000$ trajectories after thermalization, 
and sample one configuration every five trajectories, which yields $800-4000$ configurations 
for measurements.

The chiral symmetry breaking due to finite $ N_s = 16 $ can be measured in terms of 
the residual mass of the massless fermion \cite{Chen:2012jya}, 
\BAN
\label{eq:mres}
m_{res} = \frac{\left< \tr (D_c^{-1})_{0,0} \right>_U}
               {\left< \tr[\gamma_5 {D_c} \gamma_5 {D_c}]^{-1}_{0,0} \right>_U},
\EAN
where $ D_c^{-1} $ denotes the massless fermion propagator,  
``tr" denotes the trace running over the color and Dirac indices, 
and the brackets $ \left< \cdots \right>_U $ denote averaging over all configurations 
of the gauge ensemble. 
The residual masses of all gauge ensembles in this work are summarized in Table \ref{tab:mres}. 

We observe that the variation of the residual mass is quite mild, ranging from $ \sim 4.4 \times 10^{-5} $
to $ \sim 8.5 \times 10^{-5} $, i.e., less than a factor of 2.  
Moreover, the residual mass of any lattice size $ L^4 $ is much smaller 
than the energy scale $ \mu \simeq (cL)^{-1} $ of the finite-volume gradient flow scheme with $ c = 0.3 $, 
$$ (m_{res} a)_L \ll \mu a \simeq \frac{1}{c(L/a)}. $$ 
Even for the smallest $ \mu $ of the largest lattice $ 32^4 $ in this work, 
the residual mass of any gauge ensemble satisfies
$$ m_{res} a \ll  \frac{1}{0.3 \times 32 } \simeq 0.104. $$
Thus the effect of the residual masses on the renormalized couplings should be negligible for our analysis.

\begin{table}[h!]
\begin{center}
\caption{The residual masses of all gauge ensembles in this work.}
\setlength{\tabcolsep}{10pt} 
\vspace{2mm}
\begin{tabular}{ccccc}
\hline
$ 6/g_0^2 $ & $ L/a $  & $(m_{res}a)_L$ & $2L/a$  & $(m_{res}a)_{2L}$ \\
\hline
6.4650  & 16  & $ 5.8(3) \times 10^{-5} $   & 32  & $ 5.4(2) \times 10^{-5} $ \\
6.4730  & 16  & $ 5.3(1) \times 10^{-5} $   & 32  & $ 4.9(7) \times 10^{-5} $ \\
6.4750  & 16  & $ 5.7(4) \times 10^{-5} $   & 32  & $ 5.6(5) \times 10^{-5} $ \\
6.4800  & 16  & $ 5.2(8) \times 10^{-5} $   & 32  & $ 5.2(2) \times 10^{-5} $ \\
6.6000  & 16  & $ 4.6(1) \times 10^{-5} $   & 32  & $ 4.5(3) \times 10^{-5} $ \\
7.0000  & 16  & $ 4.5(6) \times 10^{-5} $   & 32  & $ 4.4(1) \times 10^{-5} $ \\
\hline
6.4610  & 12  & $ 5.9(3) \times 10^{-5} $   & 24  & $ 6.3(5) \times 10^{-5} $ \\
6.4645  & 12  & $ 5.8(2) \times 10^{-5} $   & 24  & $ 6.2(4) \times 10^{-5} $ \\
6.4680  & 12  & $ 6.2(6) \times 10^{-5} $   & 24  & $ 5.5(2) \times 10^{-5} $ \\
6.4690  & 12  & $ 5.5(3) \times 10^{-5} $   & 24  & $ 5.9(6) \times 10^{-5} $ \\
6.5700  & 12  & $ 4.9(3) \times 10^{-5} $   & 24  & $ 4.5(2) \times 10^{-5} $ \\
6.9500  & 12  & $ 4.5(7) \times 10^{-5} $   & 24  & $ 4.5(1) \times 10^{-5} $ \\
\hline
6.4590  & 10  & $ 5.8(1) \times 10^{-5} $   & 20  & $ 7.4(9) \times 10^{-5} $ \\
6.4600  & 10  & $ 6.6(5) \times 10^{-5} $   & 20  & $ 6.0(3) \times 10^{-5} $ \\
6.4640  & 10  & $ 6.5(6) \times 10^{-5} $   & 20  & $ 5.6(2) \times 10^{-5} $ \\
6.4660  & 10  & $ 6.4(3) \times 10^{-5} $   & 20  & $ 5.7(2) \times 10^{-5} $ \\
6.5500  & 10  & $ 4.8(6) \times 10^{-5} $   & 20  & $ 4.6(1) \times 10^{-5} $ \\
6.9000  & 10  & $ 4.6(6) \times 10^{-5} $   & 20  & $ 4.5(0) \times 10^{-5} $ \\
\hline
6.4490  &  8  & $ 6.2(3) \times 10^{-5} $   & 16  & $ 8.5(7) \times 10^{-5} $ \\
6.4510  &  8  & $ 6.1(4) \times 10^{-5} $   & 16  & $ 7.7(1) \times 10^{-5} $ \\
6.4520  &  8  & $ 5.9(3) \times 10^{-5} $   & 16  & $ 6.9(4) \times 10^{-5} $ \\
6.4530  &  8  & $ 6.0(2) \times 10^{-5} $   & 16  & $ 6.1(4) \times 10^{-5} $ \\
6.5200  &  8  & $ 5.1(2) \times 10^{-5} $   & 16  & $ 4.7(4) \times 10^{-5} $ \\
6.8000  &  8  & $ 4.7(6) \times 10^{-5} $   & 16  & $ 4.5(7) \times 10^{-5} $ \\
\hline
\end{tabular}
\label{tab:mres}
\end{center}
\end{table}


\section{Renormalized coupling of the finite-volume gradient flow scheme}

To obtain the renormalized coupling of gauge theory on a finite lattice with volume $ L^4 $,   
we use the finite-volume gradient flow scheme \cite{Fodor:2012td}, 
which is based on the idea of continuous smearing \cite{Narayanan:2006rf} 
or equivalently the gradient flow \cite{Luscher:2010iy}
to evaluate the expectation value $ t^2 \langle E \rangle $, 
where $ E $ is the energy density of the gauge field and $ t $ is the flow time.
This amounts to solving the discretized form of the following equation 
\BAN
\frac{d B_\mu}{dt} = D_\nu G_{\nu \mu},  
\EAN 
with the initial condition $ B_\mu |_{t=0} = A_{\mu} $, 
where $ G_{\nu\mu} = \partial_\nu B_\mu - \partial_\mu B_\nu + [B_\nu, B_\mu] $, and 
$ D_\nu G_{\nu\mu} = \partial_\nu G_{\nu\mu} + [ B_\nu, G_{\nu\mu} ] $.
As shown in Ref. \cite{Luscher:2010iy}, 
the gradient flow is a process of averaging gauge field over a spherical region of root-mean-square 
radius $ R_{rms} = \sqrt{8 t} $. Moreover, since $ t^2 \langle E \rangle $ is proportional to the 
renormalized coupling, one can use $ c = \sqrt{8t}/L $ as a constant to define a renormalization scheme 
on a finite lattice, and obtain
\bea
\label{eq:g2L}
g^2(L,a) = \frac{16 \pi^2}{3[1+\delta(c,a/L)]} \langle t^2 E(t) \rangle,  
\hspace{10mm}  E(t) = \frac{1}{2} F_{\mu\nu} F_{\mu\nu}(t),
\eea
where $ a $ is the lattice spacing depending on the bare coupling $ g_0 $, $ E $ is the energy density, 
and the numerical factor on the rhs of Eq. (\ref{eq:g2L}) is fixed 
such that $ g_c^2(L,a) = g^2_{\overline{\text{MS}}} $ to the leading order.
Here the coefficient $ \delta(c,a/L) $ includes the tree-level finite-volume 
and finite-lattice-spacing corrections \cite{Fodor:2014cpa}. 
In this paper, we use the Wilson flow, the Wilson action, and the clover observable, 
the so called WWC scheme, which is known to have very small tree-level cutoff effects \cite{Fodor:2014cpa}.
Moreover, we fix $ c = \sqrt{8t}/L = 0.30 $. 

\begin{figure}[h!]
\begin{center}
\includegraphics*[width=10cm,clip=true]{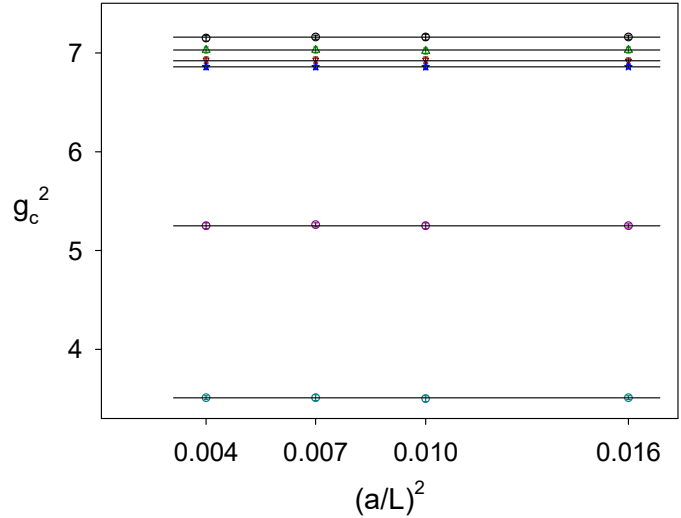} 
\caption{
Tuning $ g^2(L,a) $ on four primary lattices $ L/a=\{8, 10, 12, 16 \} $, 
for six targeted values of $ g_c^2 $. Each horizontal line is a constant fit. 
The fitted values are $ g_c^2=\{ 3.51(2), \ 5.25(2), \ 6.86(2), \ 6.92(3), \ 7.03(2), \ 7.16(2) \} $.    
}
\label{fig:g2c_tuning}
\end{center}
\end{figure}

For each targeted value of $ g_c^2 $, the renormalized couplings $g^2(L,a)$ of 
four primary lattices ($ L/a = \{ 8, 10, 12, 16 \}$) are tuned (in $ 6/g_0^2 $) 
to the same $ g_c^2 $ with statistical error less than $0.5 \% $.   
Here the statistical error is estimated using the jackknife method with a bin size of 10-15, 
of which the statistical error saturates. In Fig. \ref{fig:g2c_tuning}, we plot  
the tuned renormalized coupling $ g^2(L,a) $ versus $ (a/L)^2 $, for $ L/a = \{ 8, 10, 12, 16 \} $, 
and for six targeted values of $ g_c^2 $. Each horizontal line is a constant fit. 
The fitted values are $ g_c^2=\{ 3.51(2), \ 5.25(2), \ 6.86(2), \ 6.92(3), \ 7.03(2), \ 7.16(2) \} $,    
with $\chi^2$/d.o.f. = $\{0.19, \ 0.29, \ 0.14, \ 0.33, \ 0.19, \ 0.18 \} $, respectively. 

After the value of $ 6/g_0^2 $ is determined for a chosen $ g_c^2 = g^2(L,a) $,  
the simulation on the scaled ($s=2$) lattice is performed at the same $ 6/g_0^2 $ 
to obtain the renormalized coupling $ g^2(sL,a) $. All renormalized couplings of $ g^2(L,a) $
and $ g^2(sL,a) $ are summarized in Table \ref{tab:g2L}. 
Each row gives the values of $ g^2(L,a) $ and $ g^2(sL,a) $ at the same $ 6/g_0^2 $.
Every four rows are grouped for the same targeted value of $ g_c^2 $.

\begin{table}[h!]
\begin{center}
\caption{Summary of the renormalized couplings for all gauge ensembles in this work.}
\setlength{\tabcolsep}{10pt} 
\vspace{2mm}
\begin{tabular}{ccccc}
\hline
$ 6/g_0^2 $ & $L/a$  & $ g^2(L,a)$ & $2L/a$  & $g^2(2L,a)$ \\
\hline
6.4650	& 16 &	7.16(2)	& 32 &	7.68(3)	\\
6.4610	& 12 &	7.16(3) & 24 &  7.81(3) \\	
6.4590	& 10 &	7.16(2)	& 20 &	7.97(4) \\	
6.4490	& 8  &	7.15(3)	& 16 &	8.02(4) \\
\hline
6.4730	& 16 &	7.03(2)	& 32 &	7.53(3)	\\
6.4645	& 12 &	7.02(3)	& 24 &	7.60(3)	\\
6.4600	& 10 &	7.03(3)	& 20 &	7.67(3)	\\
6.4510	& 8  &	7.03(3)	& 16 &	7.80(2) \\
\hline
6.4750	& 16 &	6.92(3)	& 32 &	7.50(3)	\\
6.4680	& 12 &	6.93(3)	& 24 &  7.56(3)	\\
6.4640	& 10 &	6.93(3)	& 20 &  7.63(3)	\\
6.4520	& 8  &	6.93(3)	& 16 &	7.78(3)	\\
\hline
6.4800	& 16 &  6.86(2) & 32 &  7.48(3)	\\
6.4690	& 12 &  6.86(3) & 24 &  7.52(4)	\\
6.4660	& 10 &  6.86(3) & 20 &  7.61(4)	\\
6.4530	& 8  &  6.86(3) & 16 &  7.76(2)	\\
\hline
6.6000	& 16 &	5.25(2)	& 32 &	5.76(3)	\\
6.5700	& 12 &  5.25(3)	& 24 &  5.82(3)	\\
6.5500	& 10 &	5.26(2)	& 20 & 	5.88(3)	\\
6.5200	& 8  &	5.25(3)	& 16 &	5.95(3)	\\
\hline
7.0000	& 16 &  3.51(2) & 32 &  3.91(3) \\	
6.9500	& 12 &  3.50(3) & 24 &  3.97(3) \\
6.9000	& 10 &  3.51(3) & 20 &  4.01(3) \\
6.8000	& 8  &  3.51(2) & 16 &  4.14(2) \\
\hline
\end{tabular}
\label{tab:g2L}
\end{center}
\end{table}


\begin{figure}[h!]
\begin{center}
\begin{tabular}{@{}cccc@{}}
\includegraphics*[height=6cm,width=8cm,clip=true]{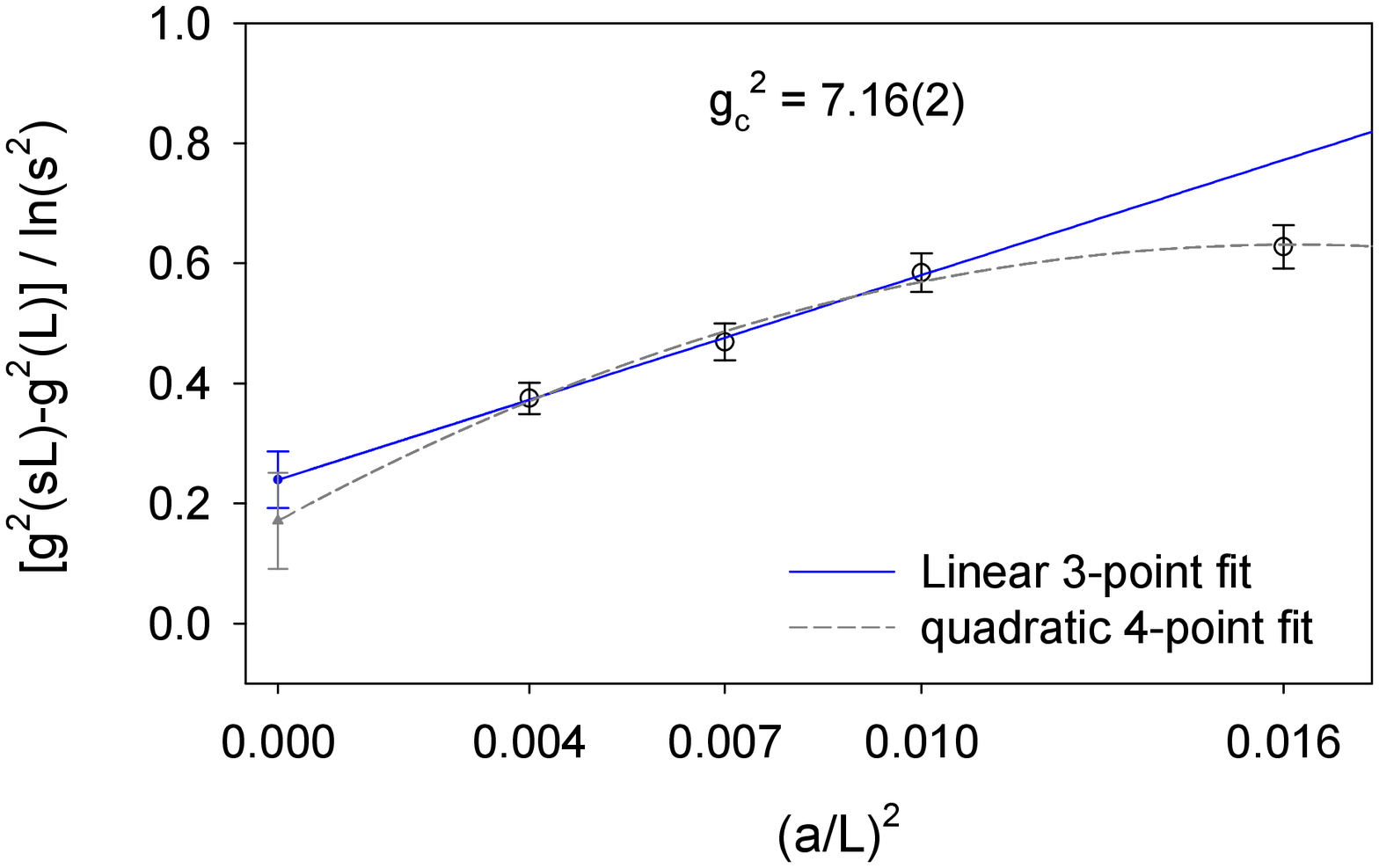}
&
\includegraphics*[height=6cm,width=8cm,clip=true]{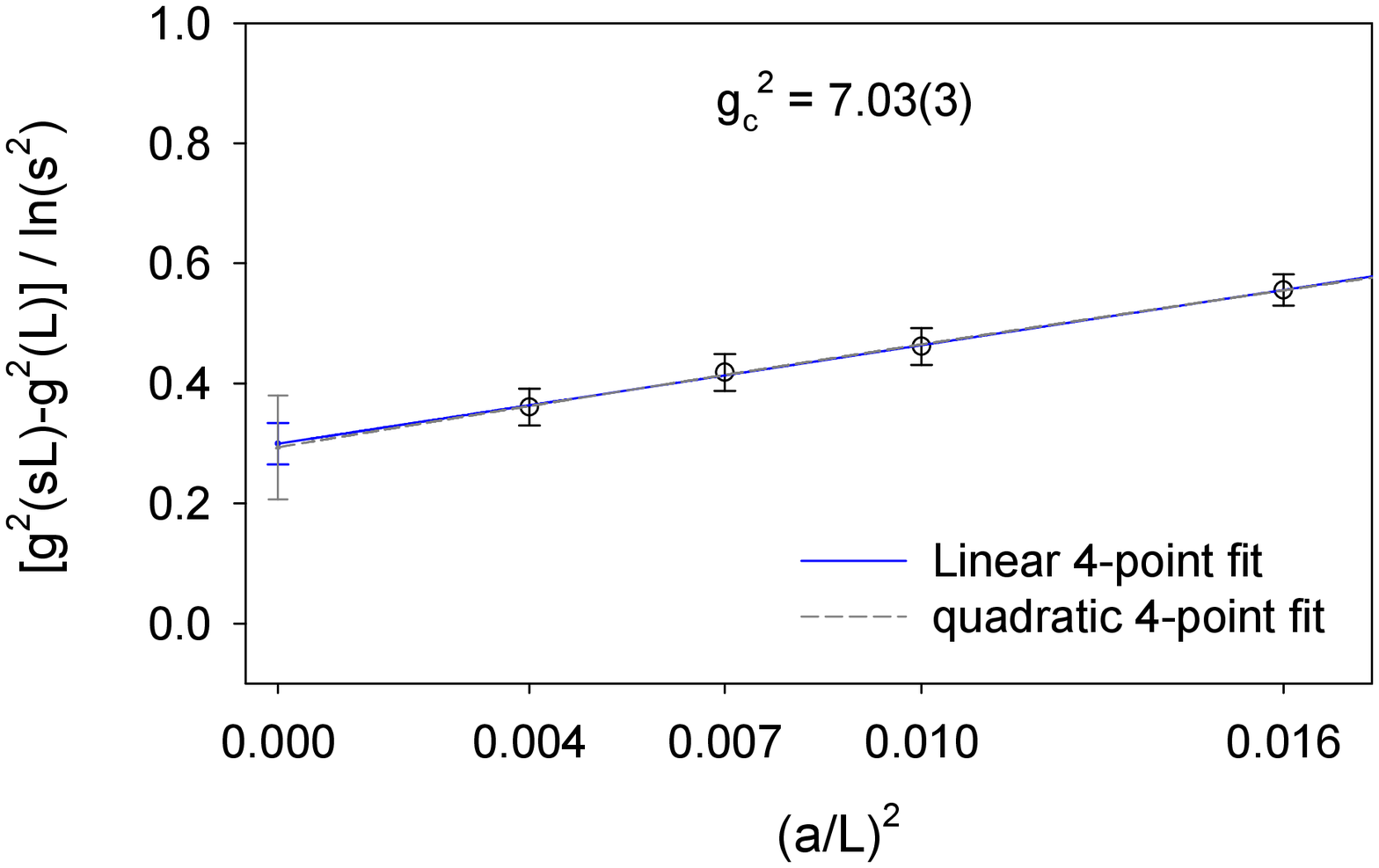} 
\\
\includegraphics*[height=6cm,width=8cm,clip=true]{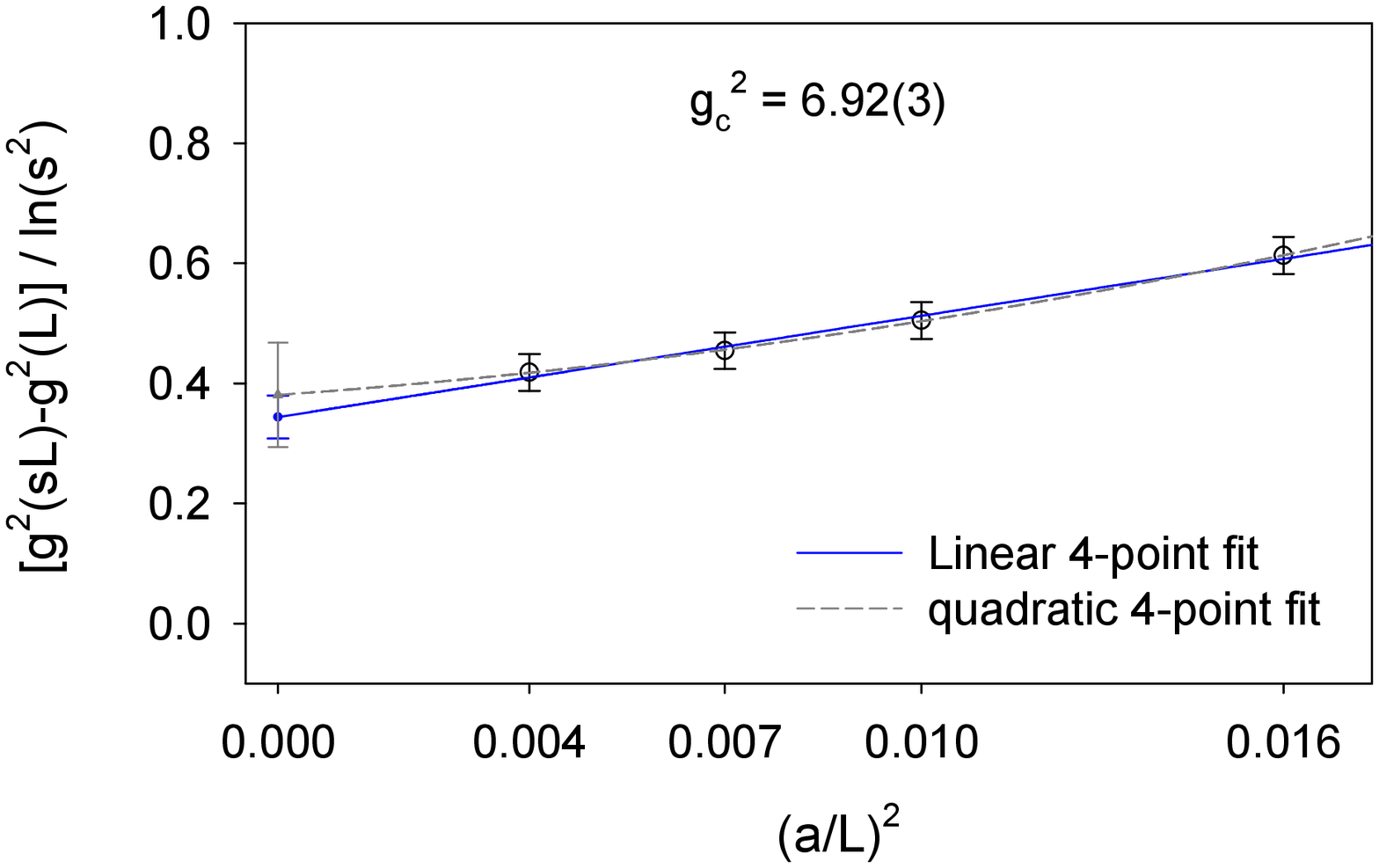}
&
\includegraphics*[height=6cm,width=8cm,clip=true]{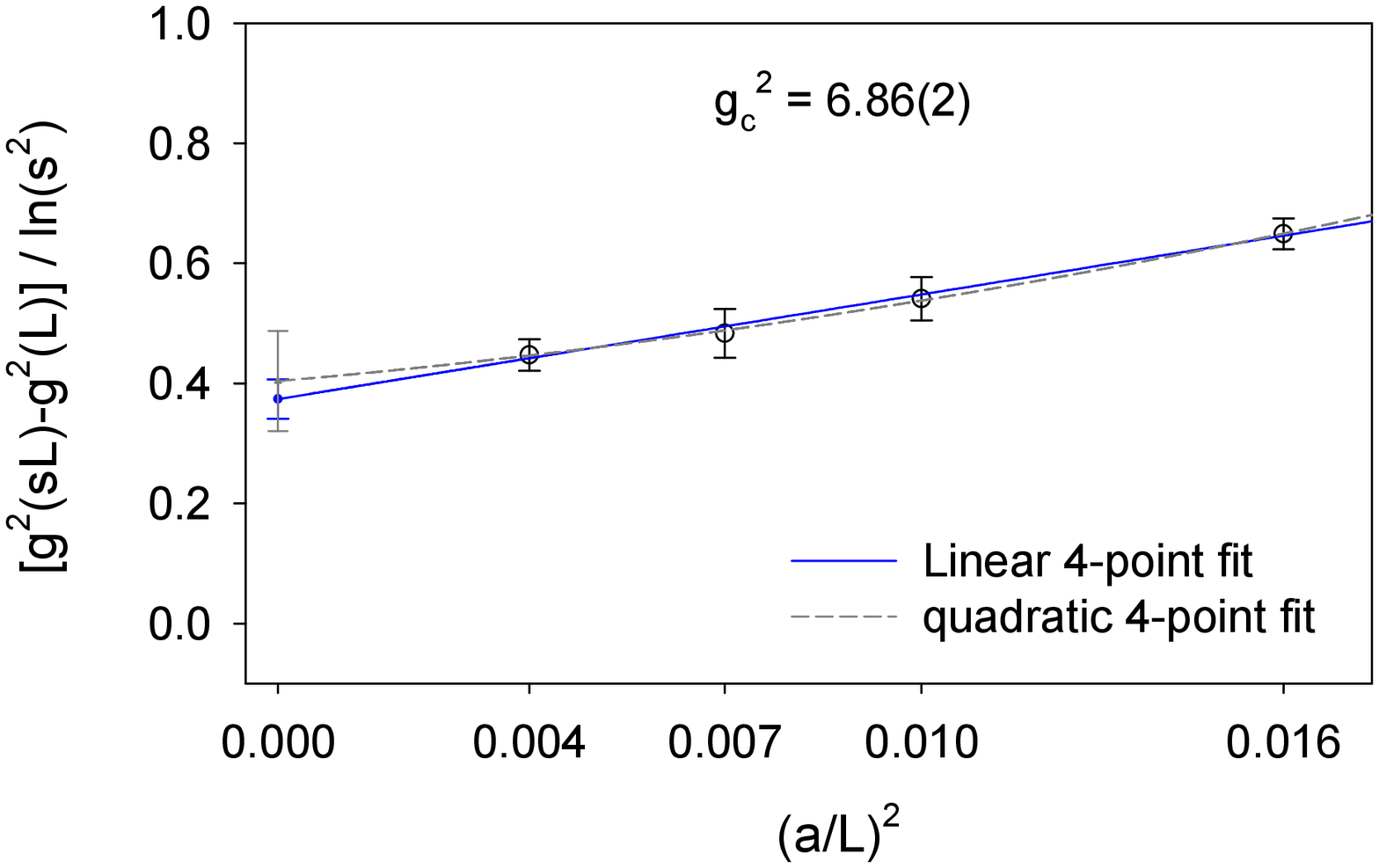} 
\\
\includegraphics*[height=6cm,width=8cm,clip=true]{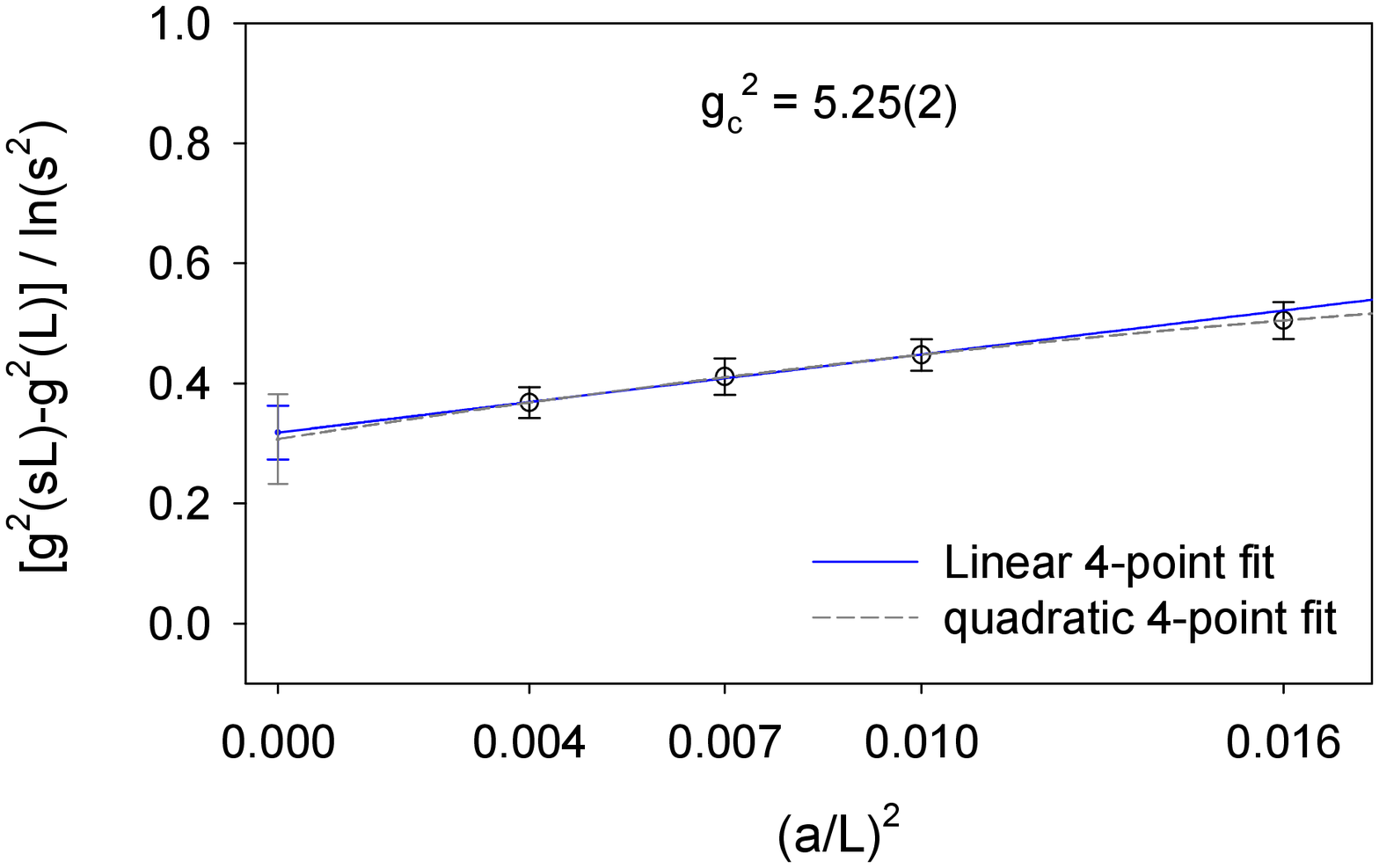}
&
\includegraphics*[height=6cm,width=8cm,clip=true]{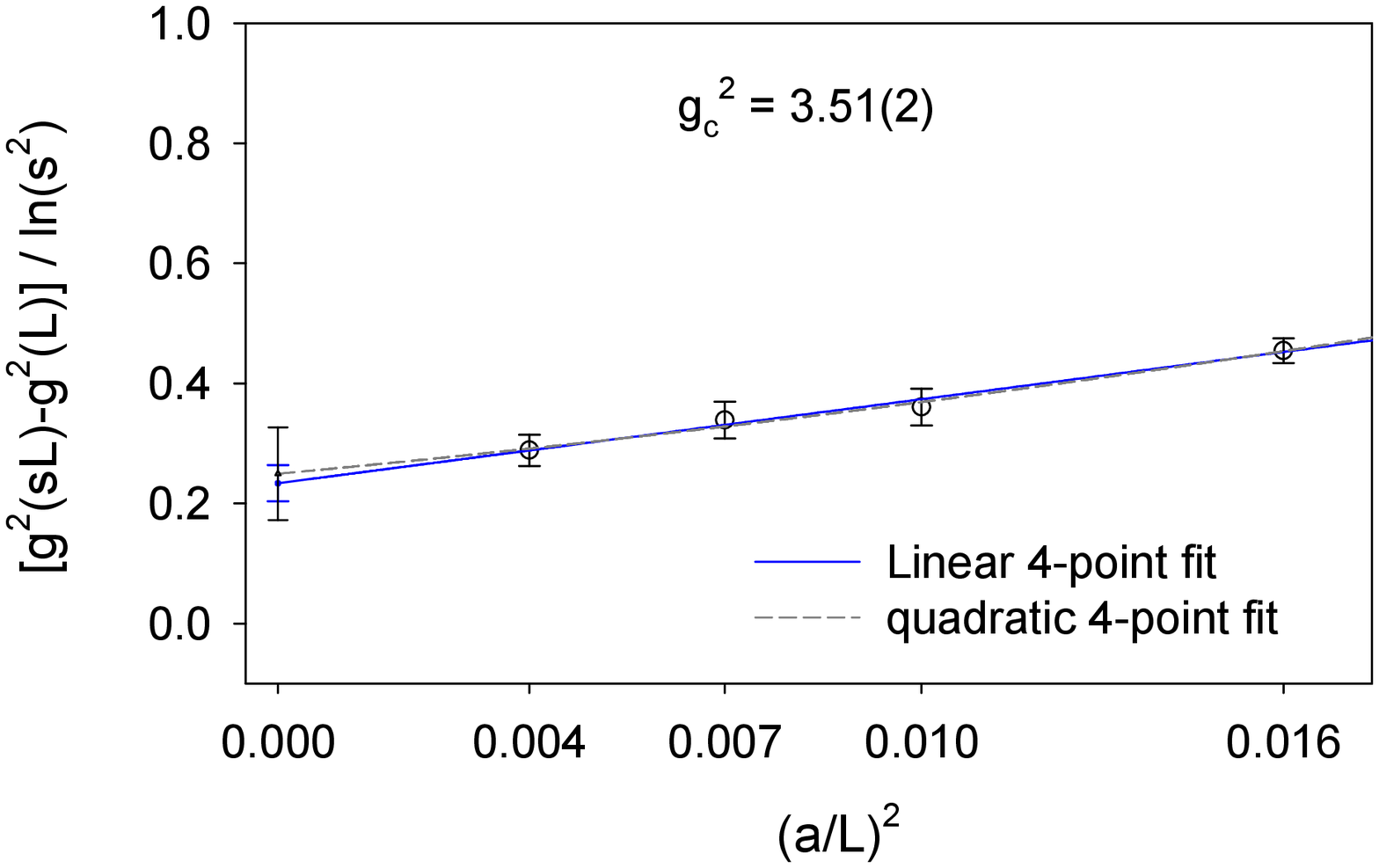}
\\
\end{tabular}
\caption{The step-scaling $\beta$-function of four lattice pairs 
$ (sL, L)/a = \{ (16,8), (20, 10), (24,12), (32, 16 \} $ are plotted versus $ (a/L)^2 $,   
for $ g_c^2 = \{ 7.16(2), \ 7.03(2), \ 6.92(3),\  6.86(2),\  5.25(2),\  3.51(2) \} $.    
The extrapolation to the continuum limit is performed with the linear fit   
and the quadratic fit, respectively. 
}
\label{fig:DBF_s2_g2c}
\end{center}
\end{figure}

\section{The step-scaling $\beta$-function $ \beta(s,a/L,g_c^2) $ and its continuum limit}

For each targeted value of $ g_c^2 = g^2(L,a) $, 
we compute the step-scaling $\beta$-function according to Eq. (\ref{eq:DBF})     
for all lattice pairs $ (sL, L)/a $ with fixed $ s=2 $. Taking the continuum limit ($a/L \to 0$),  
$ \beta(s, a/L, g^2) $ becomes $ \beta(s, g^2) $,  
\bea
\label{eq:beta_s}
\lim_{a/L \to 0} \beta(s, a/L, g^2) \equiv \beta(s, g^2)= \frac{g^2(sL) - g^2(L)}{\ln(s^2)}. 
\eea
Moreover, if $ \beta(s,g_c^2) $ is determined for several values of $ s $,  
then it can be extrapolated to $ s = 1 $,  
\bea 
\label{eq:beta_g2}
\lim_{s \to 1} \beta(s, g^2) = \beta(g^2) = \frac{d g^2(L)}{d \ln L^2}  
= -\frac{d g^2(\mu)}{d \ln \mu^2} = -\beta(g^2(\mu)),      
\eea
where $ \beta(g^2(\mu)) $ is equal to the continuum $\beta$-function in the momentum space. 
To fix our notation, we recall the $\beta$-function to two-loop order in the $ SU(3) $
gauge theory with $ N_f $ massless fermions in the fundamental representation,
\BAN
\label{eq:beta_2loop}
\beta(g^2(\mu))=\frac{d g^2}{d \ln \mu^2}=-\frac{b_1}{(4 \pi)^2} g^4 - \frac{b_2}{(4 \pi)^4} g^6 + O(g^8),
\EAN
where $ b_1 = 11 - 2 N_f/3 $, and $ b_2 = 102 - 38 N_f/3 $.

If $ \beta(g^2) $ has an IRFP, then $ \beta(s, g^2) $ also has a corresponding IRFP, and vice versa.
In this paper, we determine $ \beta(2, g^2) $ of the $ SU(3) $ 
lattice gauge theory with $N_f=10$ massless optimal domain-wall fermions in the fundamental representation, 
using four lattice pairs $ (2L, L)/a =\{ (16, 8), \ (20, 10), \ (24, 12), \ (32,16) \} $ 
for extrapolation to the continuum limit. 

In Fig. \ref{fig:DBF_s2_g2c}, $ \beta(s, a/L, g_c^2) $ is plotted versus $ (a/L)^2 $, 
for six targeted values of $ g_c^2 $.
For each targeted $ g_c^2 $, the extrapolation to the continuum limit 
($ a/L \to 0 $) is performed with the linear fit [$A + B(a/L)^2$] 
and the quadratic fit [$A + B(a/L)^2 + C (a/L)^4$], respectively.
Both fits give consistent results in the continuum limit, 
but the quadratic fits yield larger error bars.   
Note that for $ g_c^2 = 7.16(2) $, the step-scaling $\beta$-function for $ (sL, L)/a = (16,8) $
has large cutoff effects from $ (a/L)^4 $. Thus the linear fit only uses three data pounts 
from $ (sL,L)/a = \{ (20,10), \ (24, 12), \ (32, 16) \} $.   
The results for $ \beta(s,g_c^2) $ are summarized in Table \ref{tab:DBF_a0} for 
both linear and quadratic fits. 
In the following, we compare the results in the second column of Table \ref{tab:DBF_a0} 
with those of Ref. \cite{Chiu:2017kza}, and Ref. \cite{Hasenfratz:2017qyr}.

\begin{table}[h!]
\begin{center}
\caption{Extrapolation of $\beta(s, a/L, g_c^2) $ to the continuum limit.}
\setlength{\tabcolsep}{10pt} 
\vspace{2mm}
\begin{tabular}{ccccc}
\hline
$ g_c^2 $ & \multicolumn{2}{c}{linear fit} & \multicolumn{2}{c}{quadratic fit}  \\
          & $\beta(s, g_c^2)$ & $\chi^2$/dof  & $\beta(s, g_c^2) $  & $\chi^2$/dof  \\
\hline
7.16(2) &  0.239(47)  &  0.285  &  0.171(80) &  0.773 \\
7.03(2) &  0.299(35)  &  0.140  &  0.294(87) &  0.184 \\
6.92(3) &  0.344(36)  &  0.333  &  0.381(87) &  0.077 \\
6.86(2) &  0.371(32)  &  0.418  &  0.409(83) &  0.318 \\
5.25(2) &  0.318(45)  &  0.104  &  0.307(75) &  0.040 \\
3.51(2) &  0.234(30)  &  0.361  &  0.249(77) &  0.460 \\  
\hline
\end{tabular}
\label{tab:DBF_a0}
\end{center}
\end{table}

First, we check the value of $\beta(s,g_c^2) = 0.234(30) $ at $ g_c^2 = 3.51(2) $, which is   
in good agreement with the value $ 0.23(1)$ obtained in Ref. \cite{Chiu:2017kza}.
This suggests that cubic-spline interpolation can work well in the regime 
where the renormalized coupling  
varies slowly with respect to the bare coupling $ 6/g_0^2 $. 
In other words, the $\beta$-function $ \beta(s,g_c^2) $ reported in Ref. \cite{Chiu:2017kza} 
should be valid for $ 0 \le g_c^2 \le 3.51 $. 

Next, we check the value of $\beta(s,g_c^2) = 0.318(45) $ at $ g_c^2 = 5.25(2) $, which is 
quite smaller than the value $ 0.43(2) $ reported in Ref. \cite{Chiu:2017kza}.
This implies that cubic-spline interpolation fails in the regime 
where the renormalized coupling varies rapidly with respect to the bare coupling $ 6/g_0^2 $. 
Now the value of $ \beta(s,g_c^2) $ at $ g_c^2 = 5.25(2) $ is  
compatible with the result of a recent study of the $SU(3)$ gauge theory 
with $ N_f = 10 $ massless domain-wall fermions \cite{Hasenfratz:2017qyr}.
This suggests that the discrepancy between the results of Ref. \cite{Chiu:2017kza}
and Ref. \cite{Hasenfratz:2017qyr} for $ 4.5 < g_c^2 < 6.0 $ is likely due to 
the systematic error of interpolation. 

The three data points at $ g_c^2 = \{ 6.86(2), \ 6.92(3), \ 7.03(2) \} $,
$ \beta(s, g_c^2) = \{ 0.371(32), \ 0.344(36), \ 0.299(35) \}$,  
are quite larger than the corresponding ones 
$ \{ 0.06(4), \ 0.02(5), \ 0.00(8) \} $ in Ref. \cite{Chiu:2017kza}.  
This confirms that using interpolation would give unreliable results for  
$ g^2(L,a) $ and $ g^2(sL, a) $, especially in the regime where they vary rapidly 
with respect to the bare coupling $ 6/g_0^2 $, and consequently yield
an incorrect $ \beta(s,a/L,g_c^2) $ as well as the extrapolated $ \beta(s,g^2) $ in the continuum limit.
Nevertheless, the resulting $\beta(s,g_c^2) $ seems to be able to capture some salient features 
of the $\beta$-function, e.g., the increasing/decreasing trend of $ \beta(s, g_c^2) $ 
with respect to $ g_c^2$, even though it cannot give the precise shape of the entire $\beta$-function 
in the $(g_c^2, \beta)$ plane. 
  
Finally, we note that as $ g_c^2 $ is increased from 5.25(2) to 6.86(2), 
$ \beta(s,g_c^2) $ increases from 0.318(45) to 0.371(32). 
This implies that the slope of $\beta(s,g_c^2) $ is positive for 
$ g_c^2 \in [5.25, g_{max}^2] $, where $\beta(s,g_c^2) $ reaches the local maximum at $ g_{max}^2 $.  
Then for $ g_c^2 > g_{max}^2 $, the slope of $\beta(s,g_c^2) $ becomes negative, 
and $\beta(s,g_c^2) $ decreases to 0.371(32) at $ g_c^2 = 6.86(2) $. 
To determine the exact location of $g_{max}^2$ as well as 
the precise shape of $\beta(s,g_c^2)$ in the vicinity $g_{max}^2$ is very challenging, 
since it requires many targeted values of $ g_c^2 $, and also we cannot rely on 
the renormalized couplings from interpolation, especially in this regime where the slope of 
$\beta(s,g_c^2)$ changes sign (from positive to negative).

\section{Extrapolation of $ \beta(s,g^2) $}

\begin{figure}[h!]
\begin{center}
\begin{tabular}{@{}c@{}c@{}}
\includegraphics*[width=8cm,clip=true]{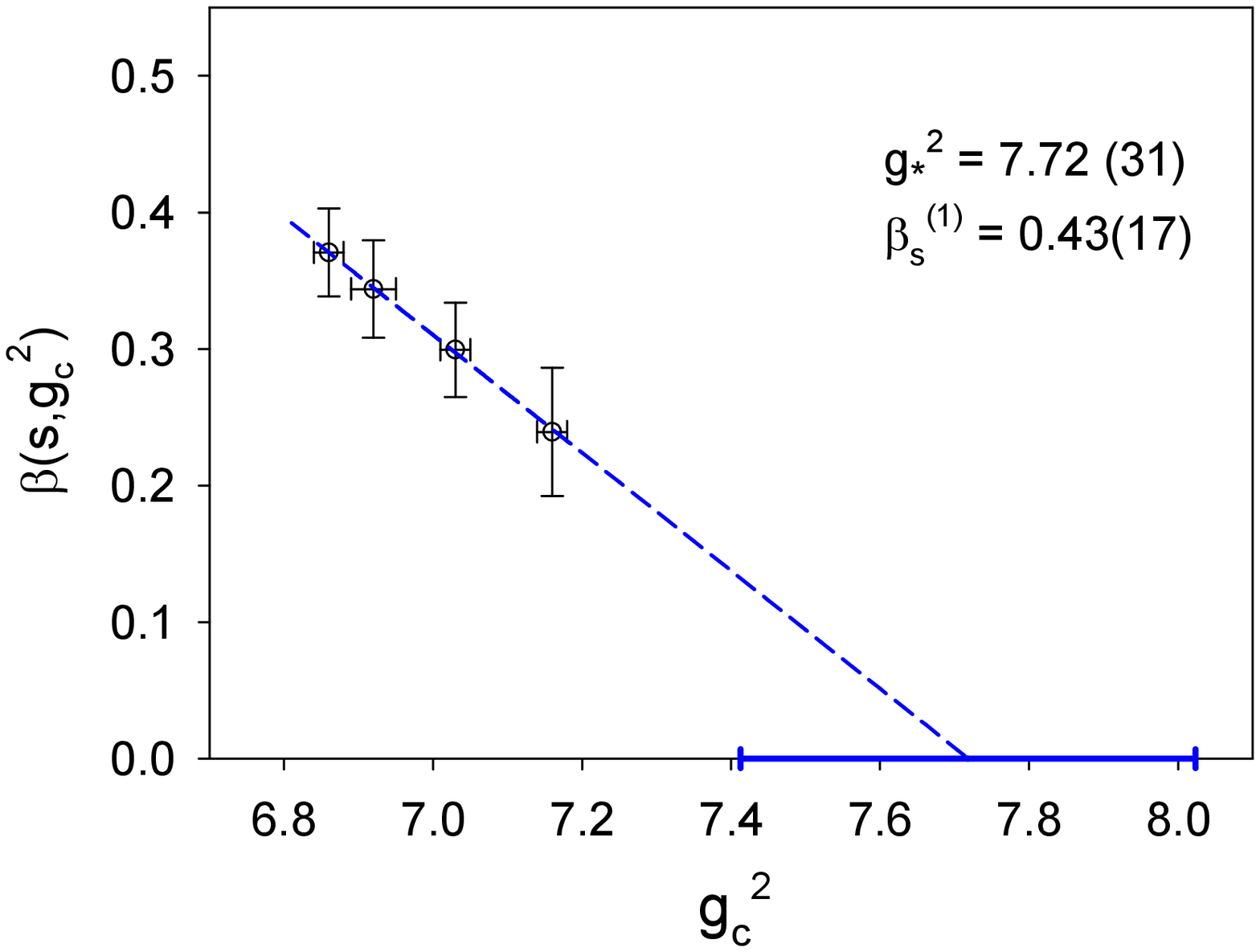}
&
\includegraphics*[width=8cm,clip=true]{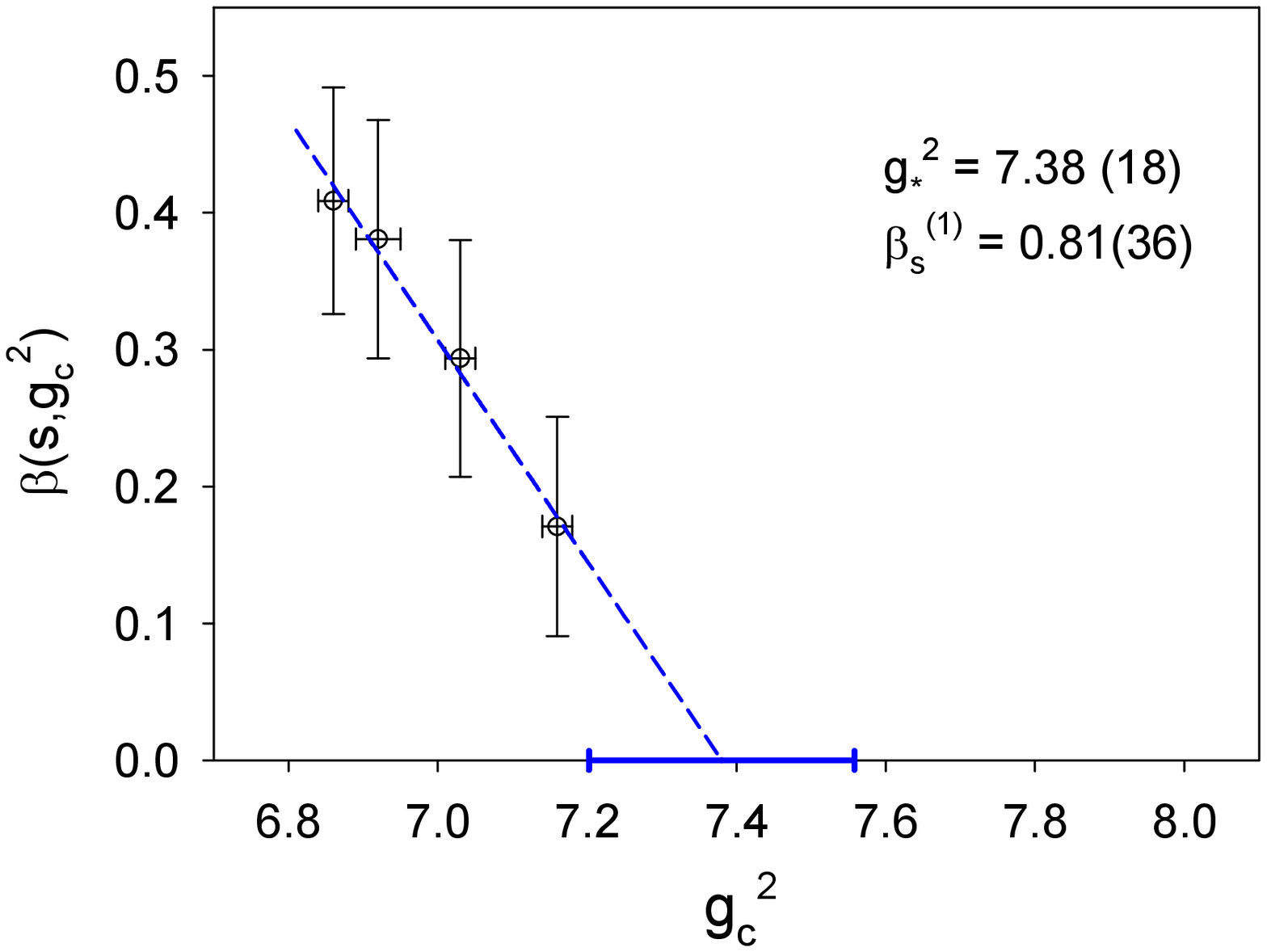}
\\ (a) & (b)
\end{tabular}
\caption{Extrapolation of $ \beta(s, g_c^2) $ with the linear fit using four data points close 
to the ``fitted IRFP" $ g_*^2 $. (a) Four data points obtained by continuum extrapolation  
with the linear fit are used (as listed in the second column of Table \ref{tab:DBF_a0}).   
(b) Four data points obtained by continuum extrapolation with the quadratic fit are used   
(as listed in the fourth column of Table \ref{tab:DBF_a0}).  
}
\label{fig:beta_g2c}
\end{center}
\end{figure}

In Fig. \ref{fig:beta_g2c}, we plot $ \beta(s,g_c^2) $ versus $ g_c^2 $,  
for $ g_c^2 = \{ 6.86(2), \ 6.92(3), \ 7.03(2), \ 7.16(2) \} $.  
In Fig. \ref{fig:beta_g2c} (a), the data points are obtained by continuum extrapolation  
with the linear fit, as listed in the second column of Table \ref{tab:DBF_a0}, 
while in Fig. \ref{fig:beta_g2c} (b), the data points are obtained by continuum extrapolation  
with the quadratic fit, as listed in the fourth column of Table \ref{tab:DBF_a0}. 
In both cases, the data points are well fitted by the linear approximation of $ \beta(s,g^2) $, 
\bea
\label{eq:beta_s_linear}
\beta(s,g^2) = \left. \frac{d \beta(s,g^2)}{d g^2} \right|_{g_*^2} (g^2-g_*^2) 
             \equiv \beta_s^{(1)} (g^2 - g_*^2).   
\eea  

In Fig. \ref{fig:beta_g2c} (a), the linear fit gives 
\bea
\label{eq:g*_a}
g_*^2 &=& 7.72 \pm 0.31, \\
\label{eq:slope_a}
\beta_s^{(1)} &=& 0.43 \pm 0.17,  
\eea
with $\chi^2$/d.o.f. = 0.06, while in Fig. \ref{fig:beta_g2c} (b), the linear fit gives 
\bea
\label{eq:g*_b}
g_*^2 &=& 7.38 \pm 0.18, \\
\label{eq:slope_b}
\beta_s^{(1)} &=& 0.81 \pm 0.36,  
\eea
with $\chi^2$/d.o.f. = 0.16.
Note that our convention for $ \beta(s,g^2) $ in Eq. (\ref{eq:beta_s}) is the negative 
of the conventional $\beta$-function in the continuum (\ref{eq:beta_g2}) and thus 
gives a negative slope $ \beta_s^{(1)} $ at the IRFP, as shown in Fig. \ref{fig:beta_g2c}.
We omit the negative sign in Eqs. (\ref{eq:slope_a}) and (\ref{eq:slope_b}) to conform with
the conventional $\beta$-function in the continuum. 
 
These two sets of results (\ref{eq:g*_a})-(\ref{eq:slope_b}) are consistent with each other 
within error bars, which seems to imply the existence of an IRFP at $ g_*^2 \in [7.20, 8.03] $.
However, we have not measured $ \beta(s, g_c^2) $ for $ g_c^2 > 7.16 $. 
Thus it is uncertain whether $ \beta(s,g_c^2) $ would behave like Eq. (\ref{eq:beta_s_linear}) 
all the way from $ g_c^2 = 7.16 $ to $ g_*^2 $, 
or if it would start to bounce back at some point $ g_{min}^2 > 7.16 $ and become
an increasing function of $ g_c^2 $ for $ g_c^2 > g_{min}^2 $. 
The former scenario implies that the theory is infrared conformal with the fixed point 
at $ g_*^2 \in [7.20, 8.03] $, while the latter suggests that the theory is near-conformal, 
depending on how closely $ \beta(s, g_c^2) $ approaches zero.

\section{Universal scaling exponent of $ \beta(g^2) $}

In the former scenario, the coefficient $ \beta_s^{(1)} $ can be used to determine
the universal scaling exponent $ \gamma_g^* $ of the $\beta$-function at the IRFP, 
\bea
\label{eq:beta_linear}
\beta(g^2) \simeq \frac{\gamma_g^*}{2} (g^2 - g_*^2), 
\eea 
with the relationship (see also Ref. \cite{Hasenfratz:2016dou})
\bea
\label{eq:gamma_g}
\gamma_g^* = \frac{\ln \left(1+ \beta_s^{(1)} \ln(s^2) \right)}{\ln (s)},  
\eea  
which can be obtained by integrating Eq. (\ref{eq:beta_g2}), and using 
Eqs. (\ref{eq:beta_linear}), (\ref{eq:beta_s}), and (\ref{eq:beta_s_linear}): 
{\small
\BAN
\ln(s^2) = \int_{L}^{sL} d \ln(L^2) = \int_{g^2(L)}^{g^2(sL)} \frac{d g^2}{\beta(g^2)}   
\simeq  \int_{g^2(L)}^{g^2(sL)} \frac{2 \ d g^2}{ \gamma_g^* (g^2 - g_*^2)} 
= \frac{2}{ \gamma_g^*} \ln \left( \frac{g^2(sL) - g_*^2}{g^2(L)- g_*^2} \right)  
\simeq \frac{2}{\gamma_g^*} \ln \left( 1 + \beta_s^{(1)} \ln(s^2) \right),   
\EAN
}
where 
\BAN
g^2(sL)=\beta(s,g^2) \ln(s^2) + g^2(L) \simeq \beta_s^{(1)}(g^2(L) - g_*^2) \ln(s^2) + g^2(L)  
\EAN
has been used.

Note that in the limit $ s \to 1 $, Eq. (\ref{eq:gamma_g}) gives 
\bea
\label{eq:gamma_g_s1}
\gamma_g^* = 2 \beta_{s=1}^{(1)} \ .
\eea
The significance of Eqs. (\ref{eq:gamma_g}) and (\ref{eq:gamma_g_s1}) is that 
the slope of $ \beta(s,g^2) $ at the IRFP (with $ s \neq 1 $) 
can be used to determine that at $ s=1 $, i.e., 
the slope of the $\beta$-function $ \beta(g^2) $ at the IRFP, which is equal to   
$ \gamma_g^*/2 $.  

Substituting Eq. (\ref{eq:slope_a}) into Eq. (\ref{eq:gamma_g}) gives   
\bea
\label{eq:gamma_g_a}
\gamma_g^* = 0.68 \pm 0.22, 
\eea 
while putting Eq. (\ref{eq:slope_b}) into Eq. (\ref{eq:gamma_g}) gives  
\bea
\label{eq:gamma_g_b}
\gamma_g^* = 1.08 \pm 0.34.
\eea 
These two results are consistent with each other within error bars. 
They are also compatible 
with the results in the weak-coupling perturbative theory, 
$ 0.473 $ (the scheme-independent value to the fifth order) and  
$ 0.853 $ (to four-loop order in the $\overline{{\text MS}} $ scheme), 
as given in Ref. \cite{Ryttov:2017kmx}. 

\begin{figure}[h!]
\begin{center}
\includegraphics*[width=8cm,clip=true]{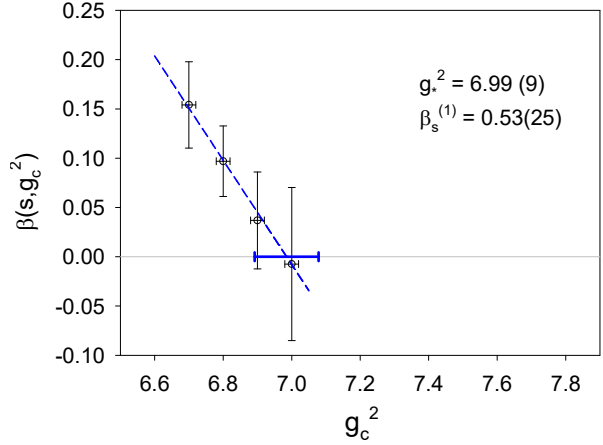}
\caption{Extrapolation of $ \beta(s, g_c^2) $ with the linear fit, using four data points  
of $ \beta(s,g_c^2) $ obtained in Ref. \cite{Chiu:2017kza}, with   
$ g^2(L,a) $ and $ g^2(sL,a) $ obtained by the cubic-spline interpolation. 
}
\label{fig:beta_g2c_inter}
\end{center}
\end{figure}

It is interesting to note that even though the interpolated $ g^2(L,a) $ and $ g^2(sL,a) $ 
in Refs. \cite{Chiu:2016uui,Chiu:2017kza}  
cannot give a reliable determination of the $\beta$-function $ \beta(s, g_c^2) $, 
especially in the regime where $ g^2(L,a) $ and $ g^2(sL,a) $ vary rapidly with respect to $ 6/g_0^2 $,  
they can still capture the slope of the $\beta$-function (at the IRFP).  
Using the four data points $ \beta(s, g_c^2) = \{ 0.154(44), \ 0.097(36), \ 0.037(49), \ -0.007(0.078) \} $ 
obtained in Ref. \cite{Chiu:2017kza} at 
$ g_c^2 = \{ 6.70(2), \ 6.80(2), \ 6.90(2), \ 7.00(2) \} $, respectively, 
the linear fit (see Fig. \ref{fig:beta_g2c_inter}) gives $ g_*^2 = 6.99(9) $ 
and the slope of the $\beta$-function $ \beta_s^{(1)} = 0.53(25) $,  
which in turn gives $ \gamma_g^* = 0.80(30) $,  
which is in good agreement with Eqs. (\ref{eq:gamma_g_a}) and (\ref{eq:gamma_g_b}).   
%
%

\section{Discussion and Conclusion}

In this paper, I performed an improved study of the $\beta$-function of $ SU(3) $ gauge theory 
with $N_f=10$ massless optimal domain-wall fermions in the fundamental representation. 
In the finite-volume gradient flow scheme with $ c = \sqrt{8t}/L = 0.3 $, 
the renormalized couplings $g^2 (L,a) $ of four primary lattices ($ L/a = \{ 8, 10, 12, 16 \}$) were 
tuned (in $ 6/g_0^2 $) to the same $ g_c^2 $ with a statistical error less than $0.5 \% $.     
Then, the renormalized couplings $ g^2(sL, a) $ of the scaled lattices ($ sL/a = \{16, 20, 24, 32\} $
with $s=2$) were computed at the same $ 6/g_0^2 $ of the corresponding primary lattices.   
Using four lattice pairs $ (sL,L)/a = \{ (16,8), (20,10), (24,12), (32,16) \} $, 
the step-scaling $\beta$-function $ \beta(a,s/L, g_c^2) $ 
was computed and extrapolated to the continuum limit $ \beta(s,g_c^2) $ 
(as summaried in Table \ref{tab:DBF_a0}) for six targeted values of $ g_c^2 $. 
Based on the four data points of $ \beta(s,g_c^2) $ at   
$ g_c^2 = \{ 6.86(2), \ 6.92(3), \ 7.03(2), \ 7.16(2) \} $ (see Fig. \ref{fig:beta_g2c}), 
two different scenarios for this theory could emerge.  

In the first scenario, $ \beta(s,g_c^2) $ would behave like Eq. (\ref{eq:beta_s_linear}) 
all the way from $ g_c^2 = 7.16 $ to $ g_*^2 $, and the theory is infrared conformal. 
Combining the fitting results from Figs. \ref{fig:beta_g2c}(a) and \ref{fig:beta_g2c}(b)  
gives $ g_*^2 = 7.55 \pm 0.36 $, and the universal scaling exponent of $ \beta(g^2) $, 
$ \gamma_g^* = 0.88 \pm 0.40 $.  

In the second scenario, $ \beta(s,g_c^2) $ would behave like a 
decreasing function of $g_c^2$ for $ g_c^2 > 7.16 $ until it reaches the local 
minimum at $ g_{min}^2 $, when it bounces back and becomes an increasing function of $ g_c^2 $ 
for $ g_c^2 > g_{min}^2 $. The question is how closely the minimum $ \beta(s,g_{min}^2) $ 
approaches zero. 

To investigate whether the theory is near-conformal or conformal for $ g_c^2 > 7.16 $ 
requires much more computing resources than that was available to this study. 
Note that the HMC simulations become more expensive as $ g_c^2 $ becomes larger 
(or, equivalently, $ 6/g_0^2 $ becomes smaller).  

Recently, a study of the $\beta $-function in the $ SU(3)$ lattice gauge theory with
$ N_f = 10 $ massless staggered fermions in the fundamental representation 
was presented in Ref. \cite{Nogradi:2018abc}, with a preview in Ref. \cite{Fodor:2017gtj}.
The continuum $\beta$-function $ \beta(s, g_c^2) $ in Ref. \cite{Nogradi:2018abc}
is a monotonic increasing function of $ g_c^2 \in [5.0, 7.7] $,  
in complete disagreement with the four data points of $ \beta(s, g_c^2) $ in Fig. \ref{fig:beta_g2c}. 
Such a dramatic discrepancy looks rather striking.

In the following, I compare the results of Ref. \cite{Nogradi:2018abc} at $ g_c^2 = 7.0 $ 
with those in this study at $ g_c^2 = 7.03(2) $.
In Ref. \cite{Nogradi:2018abc}, the step-scaling $\beta$-function $ \beta(s, a/L, g_c^2) $ 
was obtained with five lattice pairs 
$ (sL,L)/a = \{ (24,12), \ (32,16), \ (36,18), \ (40,20), \ (48,24) \} $, 
which is a monotonic decreasing function of $ (a/L)^2 $, for $ g_c^2 = 7.0 $.  
This is completely different from the $ \beta(s, a/L, g_c^2) $ in this paper, which is 
a monotonic increasing function of $ (a/L)^2 $, 
as shown in the top-right panel of Fig. \ref{fig:DBF_s2_g2c} for $ g_c^2 = 7.03(2) $.
Consequently, the continuum $\beta$-function in Ref. \cite{Nogradi:2018abc}
became very large, $ \beta(s,g_c^2) = 0.75(4) $ at $ g_c^2 = 7.0 $, which is completely different from 
the $ \beta(s,g_c^2) = 0.299(35) $ at $ g_c^2 = 7.03(2) $ in this paper (see Table \ref{tab:DBF_a0}). 
What would cause such a dramatic discrepancy between these two studies of the $\beta$-function 
of $ SU(3) $ lattice gauge theory with $ N_f = 10 $ massless lattice fermions ?  

First, could it be due to the residual mass at finite $ N_s = 16 $ in this study ?
As shown in Table \ref{tab:mres}, the residual masses are all very tiny 
and quite uniform across all lattice sizes and couplings.  
Even if the residual mass has some additive correction to the renormalized coupling, say, 
$ g^2(L,a) \to g^2(L,a) + \delta(m_{res}a) $, 
it would be canceled in the step-scaling $\beta$-function $ [g^2(sL,a) - g^2(L,a)]/\ln(s^2) $, 
since $ (m_{res} a)_{sL} \sim (m_{res}a)_{L} $. Thus the residual mass has almost
no effect on either the step-scaling function $\beta(s,a/L,g_c^2) $ or its value in the continuum limit. 
So we rule out the possibility that the residual mass
could change the slope of the step-scaling function $ \beta(s,a/L,g_c^2) $  
at $ g_c^2 = 7.03(2) $ (see the top-right panel of Fig. \ref{fig:DBF_s2_g2c}) 
from positive to negative.     
Next, does the residual mass have any effect on the shape/location of the continuum $\beta$-function 
in the $(g_c^2, \beta)$ plane ?
Now, for $ g_c^2 $ itself, $ g_c^2 \to g_c^2 + \delta(m_{res}a) $ without cancellation. 
Since $ (m_{res} a)_L $ is almost constant (with fluctuations less than $ 20\% $),  
for all $ g_c^2 $ on the primary lattices ($L/a = 8, 10, 12, 16$)  
it gives $ \delta(m_{res}a) \sim \delta $ for all $ g_c^2 $, 
and the curve of $ \beta(s, g_c^2) $ in the $(g_c^2, \beta)$ plane is shifted 
to $ \beta(s, g_c^2 + \delta ) $ with almost no change in its shape. 
If the theory is infrared conformal, the IRFP is shifted from $ g_*^2 $ to $ g_*^2 + \delta $, 
while the slope $ \beta_s^{(1)} $ of $\beta(s,g_c^2) $ at the IRFP and $ \gamma_g^* $ 
are not affected.  In view of the tiny residual masses in Table \ref{tab:mres},  
I suspect that $ \delta $ is already much smaller than the error 
of $ g_c^2 $ resulting from tuning $ g^2(L, a) = g_c^2 $ for all primary lattices. 
In general, for any study with DWFs, if $ \delta(m_{res}a) $ is a monotonically increasing function 
of $ g^2(L,a) $, then the shape of the curve $ \beta(s,g_c^2) $ would be a little bit stretched 
along the positive direction of the $g_c^2 $ axis, due to the non-uniformity of $ \delta(m_{res}a) $.
If the theory is infrared conformal, the measured location of the IRFP would be 
a little larger than the exact $ g_*^2 $ (at zero residual mass), 
and also the measured slope of the $\beta$-function at the IRFP 
would be smaller than its exact $ \beta_s^{(1)} $.  
Consequently, the measured universal scaling exponent would be a little smaller than 
the exact $ \gamma_g^* $ (at zero residual mass). 
Likewise, if the theory is infrared near-conformal,    
the measured $ g_{min}^2 $ would be a little larger than 
the exact $ g_{min}^2 $ (at zero residual mass). 
From the above discussions, the effect of the residual mass in this study should be very small. 
Thus it is impossible to change the slope/curvature of $ \beta(s,g_c^2) $ 
in Fig. \ref{fig:beta_g2c} from negative to positive. 
So we rule out the possibility that the residual mass could produce such a dramatic discrepancy  
in $ \beta(s, g_c^2) $ at $ g_c^2 \sim 7.0 $, namely,   
$ 0.299(35) $ in Table \ref{tab:DBF_a0} versus $ 0.75(4) $ in Ref. \cite{Nogradi:2018abc}.

Next, could this be due to the volumes being too small in this study ?
Would it be possible to make a dramatic change in the continuum extrapolation  
if we include a larger volume, say, $48^4$ in our analysis ? 
From the data for $ \beta(s, a/L, g_c^2) $ at $ g_c^2 = 7.03(2) $,  
as shown in the top-right panel of Fig. \ref{fig:DBF_s2_g2c}, 
the rate of change of $ \beta(s, a/L, g_c^2) $ with respect to $ (a/L)^2 $ 
is rather small at any $ (a/L)^2 $. 
Even if we add a larger volume, say $ 48^4 $, with an additional data point of  
$ \beta(s,a/L,g_c^2) $ at $(a/L)^2 =(1/24)^2 \sim 0.00174 $,  
it is very unlikely that the slope of $ \beta(s, a/L, g_c^2) $
would undergo a dramatic change from a small positive slope to a large negative slope in the 
limit $(a/L) \to 0 $.
Note that the $ (a/L)^4 $ correction gets smaller for larger $ L $ as $ (a/L) \to 0 $. 
Consequently, the deviation of the step-scaling $\beta$-function $\beta(s, a/L, g_c^2) $ 
from the linear function of $ (a/L)^2 $ gets smaller as $ L $ gets larger. 
In other words, in this study, adding an extra data point with a larger volume 
for the step-scaling function $ \beta(s,a/L,g_c^2) $ at $ g_c^2 = 7.03(2) $  
is very unlikely to make a dramatic change to its value in the continuum limit ($ a/L \to 0 $). 
Note that in this study only one data point of $\beta(s,a/L,g_c^2)$ at the largest $ g_c^2 = 7.16(2) $ 
and at the smallest volume with $ (a/L)^2 = (1/8)^2 \sim 0.016 $ has a noticeable correction 
from the $ (a/L)^4 $ term, as shown in the top-left panel of Fig. \ref{fig:DBF_s2_g2c}.    
On the other hand, if we omit the data point of the largest volume with $ (a/L)^2 = (1/16)^2 \simeq 0.004$
in the top-right panel of Fig. \ref{fig:DBF_s2_g2c} for $ g_c^2 = 7.03(2) $,
and perform the continuum extrapolation with the linear fit, we get $ \beta(s, g_c^2) = 0.306(54) $
with $ \chi^2$/d.o.f. = 0.13, which is in good agreement with the result $ 0.299(35) $ obtained
with four lattice pairs, as given in Table \ref{tab:DBF_a0}.
Thus we rule out the possibility that adding data points of $ \beta(s,a/L,g_c^2) $ with larger volumes 
in this study could produce such a dramatic difference 
in $ \beta(s, g_c^2) $ at $ g_c^2 \sim 7.0 $, 
namely, $ \sim 0.3 $ in Table \ref{tab:DBF_a0} versus $ \sim 0.75 $ in Ref. \cite{Nogradi:2018abc}.
 
Finally, we compare the actions in this study with those in Ref. \cite{Nogradi:2018abc}.
The gauge action in Ref. \cite{Nogradi:2018abc} is the tree-level improved Symanzik gauge action, 
which is different from the Wilson plaquette action in this study. However, we do not expect that 
different gauge actions would cause such dramatic differences in any observables.
Then we come to the possibility that the dramatic discrepancies 
are due to two different lattice fermion actions. 
If both lattice fermion Dirac operators belong to the same universality class 
of the continuum Dirac operator, then they should produce consistent results in the continuum limit. 
Could the staggered fermion operator violate fermion universality in the vicinity of the IRFP ? 
This conjecture has been addressed by the authors of Ref. \cite{Hasenfratz:2017qyr}; 
however, it was refuted by the authors of Ref. \cite{Fodor:2017gtj}. 
A nonperturbative analytic proof seems to be required to settle the issue 
of whether the (rooted) staggered fermions belong to the   
same universality class of the continuum Dirac operator, especially in the vicinity of the IRFP.
At the moment, the results of this study could not rule out those in Ref. \cite{Nogradi:2018abc}, 
and vice versa. 
Moreover, I do not see any other (systematic/statistical) possibilities that can reconcile 
the dramatic discrepancies between these two studies of the $\beta$-function 
of the $ SU(3) $ lattice gauge theory with $ N_f = 10 $ massless lattice fermions.  

To conclude, based on the four data points of $ \beta(s,g_c^2) $    
as shown in Fig. \ref{fig:beta_g2c}, 
I infer that the theory is infrared near-conformal, or conformal with the fixed-point $ g_*^2 = 7.55(36) $.
This also implies that the $ SU(3) $ 
gauge theory with $ N_f = 12 $ massless fermions in the fundamental representation  
is most likely infrared conformal with IRFP $ g_*^2 < 7.2 $. This prediction is consistent with 
a recent study with $ N_f = 12 $ domain-wall fermions \cite{Hasenfratz:2018wpq}, 
which suggests that the theory is infrared conformal with an IRFP $ g_*^2 \sim 6 $.

\begin{acknowledgments}

This work is supported by the Ministry of Science and Technology 
(Grant Nos.~107-2119-M-003-008, 105-2112-M-002-016, 102-2112-M-002-019-MY3), 
and the National Center for Theoretical Sciences (Physics Division).
All computations were performed on GPU clusters at Academia Sinica Grid Computing Center (ASGC), 
Information Technology Center of National Taiwan Normal University, 
and Physics Department of National Taiwan University. 
We gratefully acknowledge the computer resources and the technical support provided by these institutions. 
   
\end{acknowledgments}


\end{document}